\documentclass[12pt]{article}
\begin{document}

\input epsf
\def\figplot#1#2#3#4{\begin{figure}[htbp] \centering \leavevmode
\epsfxsize=#4\columnwidth \epsfbox{#1} \caption{#2} \label{#3} \end{figure}}

\def\twopanelplot#1#2#3#4#5{\begin{figure}[htbp] \centering \leavevmode
\epsfxsize=#5\columnwidth \epsfbox{#1} \vspace{\medskipamount}
\epsfxsize=#5\columnwidth \epsfbox{#2} \caption{#3} \label{#4} \end{figure}}

\def\etal   {{\sl et~al.}}
\def\wisk#1{\ifmmode{#1}\else{$#1$}\fi}
\def\um     {\wisk{{\rm \mu m}}}
\def\arcmin   {\wisk{^\prime\ }}
\def\arcsec   {\wisk{^{\prime\prime}\ }}
\def\gsim   {\wisk{_>\atop^{\sim}}}

\def\lsim   {\wisk{_<\atop^{\sim}}}

\def\lsun   {\wisk{{\rm L_\odot}}}

\title{THE SUBMILLIMETER FRONTIER: A SPACE SCIENCE IMPERATIVE}
\author{
John C. Mather,$^1$ 
S. Harvey Moseley, Jr.,$^1$ 
David Leisawitz,$^1$\\
Eli Dwek,$^1$ 
Perry Hacking,$^2$ 
Martin Harwit,$^3$ 
Lee G. Mundy, $^4$\\
Richard F. Mushotzky,$^1$ 
David Neufeld,$^5$ 
David Spergel,$^6$ 
Edward L. Wright$^7$\\ \\ 
{\normalsize \em $^1$ NASA Goddard Space Flight Center}\\
{\normalsize \em $^2$ Vanguard Research}\\ 
{\normalsize \em $^3$ 511 H St., Washington, DC; also Cornell University}\\
{\normalsize \em $^4$ University of Maryland}\\
{\normalsize \em $^5$ Johns Hopkins University}\\
{\normalsize \em $^6$ Princeton University}\\
{\normalsize \em $^7$ UCLA}
}

\maketitle

\section*{Abstract}

A major goal of modern astrophysics is to understand the processes by which the
universe evolved from its initial simplicity, as seen in measurements of the
Cosmic Microwave Background (CMB), to the universe we see today, with complexity
on all scales.  The initial collapse of the subtle seeds seen in the CMB results
in the formation of galaxies and stars.  The formation of the first stars marks
the beginning of heavy element nucleosynthesis in the universe, which has a
profound effect on the formation of subsequent generations of stars.  The
density fluctuations in the CMB from which all this structure grows are
primordial.  The development of these fluctuations into progressively more
complex systems is the history of the universe; galaxies from seed structures,
clouds from uniform interstellar gas, stars from clouds, elements from the
initial hydrogen and helium, molecules from elements, dust from molecules, and
planets from dust.  These processes have resulted, at least on Earth, in the
remarkable range of physical, chemical, and biological systems we see around us.

While the diffuse background measurements of COBE reveal the importance of the
far infrared and submillimeter in early galaxy and star formation, the
understanding of the development of complex structure requires high resolution
imaging and spectroscopy.  It is clear from COBE observations of the far
infrared background that much of the luminosity emitted in the critical initial
phases of structure formation is emitted at far infrared wavelengths.  Dust,
which is responsible for the far infrared emission, hides much of the activity
in the early universe from optical and near infrared study.

We present a concept for an instrument that will enable us to observe the
dominant far infrared and submillimeter emission from the epoch of
structure formation.  The instrument, called the {\bf Submillimeter Probe of the
Evolution of Cosmic Structure (SPECS)}, will produce high resolution images
and spectroscopic data, allowing us our
first clear view into the hidden environments where the structures in the
universe developed.  These observations will become powerful new tools
to understand this crucial phase in the development of complex structures in the
universe.  SPECS will also open up new realms of discovery in the local
universe.  For example, individual star forming regions in a wide variety of
nearby galaxies can be studied in detail, showing how the star formation process
is affected by variable conditions such as metallicity and interstellar
radiation field intensity.  Such information will further improve our
understanding of
galaxies in the early universe.

SPECS is not only scientifically exciting, it is technically feasible.  If
a concerted effort is made to advance and test the required technologies during
the next decade, it will be possible to build the SPECS observatory in the
succeeding years.

\section{Introduction}
\label{sec:intro}

One of the most remarkable understandings developed by modern astrophysics is
that the very complex local universe evolved from an earlier hot phase of almost
perfect uniformity.  Observations of the CMB reveal density fluctuations of only
a few parts in $10^5$ at $z\sim 1000$.  By the time we see the distant universe
of galaxies in the Hubble Deep Field $(z\sim 1 - 3)$, these density fluctuations
have developed into isolated galaxies and clusters in mostly empty space, with a
whole array of shapes and sizes.  {\bf The primary goal of SPECS is to provide a
definitive observational basis for understanding the history of and the
processes that drive the development of complex structure from the homogeneous
early universe.}  While simple observations are sufficient to characterize
simple systems, the rich complexity of this era of galaxy formation requires
observations with sufficient spatial and spectral resolution to characterize the
total luminosities, physical conditions, and morphological characteristics of
these developing systems.

\begin{table}
\begin{centering}
\begin{tabular}{|l|}
\hline
{\bf Formation of stars from primordial material and }\\
\hspace*{0.5cm}{\bf galaxies from pre-galactic structures}\\
{\bf Evolution of galaxies and structures}\\
{\bf History of energy release, nucleosynthesis, and dust formation}\\
{\bf Feedback effects of Population III stars on those that came later}\\
\hline
\end{tabular}
\caption{SPECS will enable detailed astrophysical studies of the early universe.}
\label{tbl:goals}
\end{centering}
\end{table}

Observations with HST, ISO, Keck, JCMT and other large Earthbound telescopes
have resulted in the identification of a large number of galaxies at redshifts
out to $z>3$, and have begun to produce a consistent picture of cosmic
evolution.  The optically selected samples of galaxies show a rapid increase in
star formation rate, hence luminosity, as a function of redshift, peaking at
about 10 times that in the local universe at $z \sim 1.5$ (Madau, Pozzetti \&
Dickinson 1998).  The far-infrared/submillimeter extragalactic background
measured by the DIRBE and FIRAS instruments on COBE (Hauser \etal\ 1998; Fixsen
\etal\ 1998) requires a similarly high rate of star formation at $z\sim 1.5$
(Dwek \etal\ 1998).

Conclusions regarding the earlier history of star formation are more tentative;
the star formation rate may remain near the elevated rate seen at $z \sim 1.5$,
or it may decline significantly.  The background measurements provide a weak
constraint on the star formation at $z \gsim 1.5$.  Ground based submillimeter
observations from the SCUBA camera on the JCMT and measurements with ISOCAM on
ISO have recently revealed high z galaxies with very high star formation rates.
These galaxies are completely undistinguished in an optical survey.
Observations of the Hubble Deep Field at 450 and 850 $\mu$m suggest that
optically faint galaxies undergoing massive starbursts may be responsible for
50\% of the cosmic infrared background seen by FIRAS and DIRBE, and that as much
as 80\% of the luminosity of the early universe may be emitted in the far
infrared (Hughes \etal\ 1998).  Apparently most of the nucleosynthesis and its
associated energy release, the motive force behind galactic evolution, occurs in
environments that in optical studies are shrouded in dust.  Indeed, extinction
by dust is an important and ill-quantified source of systematic error in many
cosmological surveys.

{\bf Thus we should strive to obtain a detailed view of the optically obscured
star forming systems in the early universe; we need the ability to measure the
luminosities, redshifts, metal abundances and morphologies of galaxies back to
the epoch of their formation.  SPECS achieves this goal with sensitive
submillimeter interferometry and spectroscopy.}  As outlined in
Table~\ref{tbl:parameters}, SPECS provides high sensitivity and HST-like angular
resolution in the far infrared, a wide field of view, and spectral resolution
$\sim 10^4$.  Since submillimeter radiation from the early universe is faint,
cryogenic telescopes with background-limited direct detectors are required.  The
angular scales of the relevant structures are very small, so interferometric
baselines ranging up to $\sim1$ km are required.

\begin{table}
\begin{centering}
\begin{tabular}{|l|l|}
\hline
{\bf Telescopes}		& {\bf Three, $D_t = 3$ m aperture}\\
{\bf Telescope temperature}	& {\bf 4 K}\\
{\bf Maximum baseline, $b_{max}$}	& {\bf up to 1 km}\\
{\bf Detectors}		& {\bf Six $N_{\rm pix} \times N_{\rm pix} = 100 \times 100$ detector arrays} \\
{\bf Detector type}		& {\bf Superconducting Tunnel Junction or bolometer} \\
{\bf Spectrometer}		& {\bf Michelson Interferometer} \\
{\bf Wavelength range}	& {\bf 40 - 500 $\mu$m} \\
{\bf Spectral resolution}	& {\bf up to 10$^4$} \\
{\bf Angular resolution}	& {\bf $\lambda/b_{max}$, 0.05\arcsec\ for 250 \um\ and 1 km} \\
{\bf Field of view}		& {\bf $N_{\rm pix} \frac{\lambda}{2D_t}$, 14\arcmin\ for 250 \um, $N_{\rm pix} = 100$, Nyquist sampling}\\
{\bf Typical image size}	& {\bf $\sim 17000 \times 17000$ resolution elements} \\
{\bf Typical exposure}	& {\bf 1 day} \\
{\bf Typical sensitivity}	& {\bf $\nu S_\nu$ 10$^7$ Hz-Jy, 10$^{-19}$ W/m$^2$ at 100 \um, 1 $\sigma$}\\
\hline
\end{tabular}
\caption{Parameters of SPECS, Submillimeter Probe of the Evolution of Cosmic Structure}
\label{tbl:parameters}
\end{centering}
\end{table}

Here we present a program of development that can, by late in the next decade,
open the era of galaxy formation to detailed study.  The emission from stars
from this era is redshifted into the near infrared spectral region, where it can
be studied by NGST.  When combined with NGST observations, SPECS observations
will provide direct measures of the total luminosities of forming and young
galaxies by measuring their fluxes in the two spectral regions where the bulk of
their energy is emitted.

SPECS will help answer several key questions about the basic characteristics of the
universe:
\begin{enumerate}
\item When was ``first light''?  Did the first generation of stars form in early
galaxies or before such systems existed?

\item What is the history of energy release and nucleosynthesis in the universe?
How did carbon, oxygen, other heavy elements, and dust build up over time?  What
mechanisms were responsible for dispersing the metals?

\item Did the process or rate of star formation change over the course of cosmic
history?  How might any change in the star formation process be attributed to
the gradual enrichment of the interstellar medium with heavy elements, or other
factors still unknown?

\item What are the processes of structure formation in the universe?  Were these
processes hierarchical?  What is the role of collisions between clouds and
galaxy fragments?  When and how did the first bulges, spheroids and disks form?
How, ultimately, did the galaxies in today's universe form?
\end{enumerate}

In addition to these primary objectives, SPECS will provide unprecedented
observations needed to understand the formation of stars and planets and the
interaction of stars, at birth and death, with the interstellar medium.

In this paper, we argue for the necessity of a sensitive submillimeter
interferometer for spectroscopic and imaging studies of the development of
structure in the universe and the enrichment of the universe with heavy elements
over cosmic time.  Section~\ref{sec:context} provides a theoretical and
observational context in which gaps in our understanding are evident and the
niches to be filled by SPECS are identified.  Section~\ref{sec:physproc} relates
the physical processes that yield far-IR and submillimeter emission to the astrophysical
systems that produce the emission, and illustrates how the SPECS observations
can be used to address the scientific questions mentioned above.  A technical
concept for the SPECS observatory is presented in \S \ref{sec:SPECS}.  We show
that many of the technologies required for its realization exist, or are being
developed as a part of the NASA program (\S \ref{sec:todaytomorrow}).  We
identify the new technologies required (\S \ref{sec:technology}), and outline an
incremental scientific program in which they can be developed during the next
decade (\S \ref{sec:roadmap}).

\section{Cosmic Evolution - Current Understandings and \\ \mbox{Observational} Context}
\label{sec:context}

The canonical picture of the evolution of the universe given below provides a
framework for discussion of the need for an instrument like SPECS.  As they are
currently understood, the major developments were as follows:

\begin{itemize}
\item $z >> 10^{7}$ -- The expanding universe begins in a hot, dense Big
Bang, including a period of cosmic inflation that produced a smooth
distribution of matter over the scale of our horizon and the density
fluctuations, a part in 10$^5$, seen in the cosmic microwave background
radiation.  A few minutes later, nucleosynthesis results in the production of
H, D, He, Li, Be, and B from the initial protons and neutrons.  Observed
abundances are consistent with this picture.  Dark matter, perhaps both cold and
hot, begins to move under its own gravity as soon as the distant universe came
within the causal horizon.

\item $z \sim 1000$ -- The decoupling of radiation from matter allows
baryonic matter to cluster around the dark matter.  The universe becomes
transparent, leaving behind the microwave background radiation observed by COBE.
We know the statistics of the density field at this time, and believe that a
linear theory describes the growth of density fluctuations.  This is the basis
of the claim that the cosmic microwave background fluctuations can measure the
main cosmic parameters.  The MAP (2000) and Planck (2007) missions, and many
ground based measurements, will do this.

\item $z \sim 20$ -- The first luminous objects form in fluctuations of greatest
density enhancement ($ > 3\sigma$).  They must cool off as they collapse,
emitting spectral lines from H$_2$ and H.  Such lines would be extremely weak
(of the order of pico-Janskys) and are not expected to be observable.  However,
the first star-forming systems might be detectable; whereas the binding energy,
only $\sim 1$ -- 10 eV/nucleon, is released during cloud collapse, nuclear
fusion releases a few MeV per nucleon.  In this epoch massive stars (Population
III?)  form and begin to ionize the intergalactic medium, and some expel heavy
elements in supernova explosions and stellar winds.  These objects are entirely
hypothetical, but we know that something ionized the intergalactic medium by a
redshift of 5.  {\bf With SPECS we will be able to find and characterize the
first stars even if our view is obscured by dust, learn when these objects
produced heavy elements, and determine when the first dust formed.}  At a
redshift of 20, Ly$\alpha$ is observable by NGST.  The redshifted Brackett lines
could be observed by SPECS, as could the near IR emission from cool stars,
continuum emission from dust in H~II regions, and several important diagnostic
fine structure lines from heavy elements (see \S \ref{sec:physproc}).  The [C~II]
158 \um\ line is redshifted to 3.3 mm and could be seen with the Millimeter
Array (MMA), along with thermal emission from dust associated with the neutral
phases of the ISM.

\item $z \sim 3 - 20$ -- Secondary structure formation.  Cloud cooling is
enhanced by the inclusion of newly synthesized heavy elements.  Galaxies grow by
collisions and absorption of smaller fragments, with a rate governed by the
statistics of the primordial density fluctuations and their growth.  Many are
very dusty, with star formation obscured by very local dust from young hot stars
and supernovae.  Interstellar shocks reprocess the dust.  Some heavy elements
enrich the newly ionized intergalactic medium, driven by high pressures and
outflows from small galaxies with insufficient gravity to retain the debris.
Gas liberated from the galaxies by collisions is heated and radiates X-rays.
Deep potential wells form in clusters and gravitational lenses allow far IR
measurements of even more distant objects.  SIRTF will count the galaxies and
measure their luminosity functions.  NGST will observe the stars in the
galaxies, and the obscuration by dust.  {\bf SPECS will observe the dust
luminosity directly, allowing the inference of the hidden stellar luminosity.}

\item $z \sim 1 - 3$ -- Star formation peaks with frequent collisions of
galaxies and fragments.  Cooling flows allow hot gas to fall to the centers of
galaxy clusters and disappear from view, possibly forming stars (O'Connell and
McNamara 1991).  In many galaxies Active Galactic Nuclei (AGN) form in dense
cores with accretion disks of infalling material and produce jets observable
from radio to X-ray.  These AGN have strong effects on their host galaxies.
SIRTF establishes the relationship between AGN and the ultraluminous infrared
galaxies first seen by IRAS.  HST reveals the morphologies and colors of young
galaxies and tells us that collisions are common.  NGST sees the stars and,
depending on its long wavelength coverage, part way into the dense cores of AGN.
SPECS sees into the dense cores, penetrating the opaque dust and seeing its
emission.  A number of H$_2$O lines, which are important diagnostics of star
forming molecular clouds (Harwit \etal\ 1998), will be visible to SPECS.  {\bf
The combination of NGST and SPECS allows us to distinguish the emission from
individual star forming regions and learn whether and how much star formation is
driven by collisions of galaxies, spiral density waves in existing galaxies,
feedback from other star formation, or other causes.}

\item $z \sim 0,$ the local universe -- Galaxy mergers have nearly ceased.
Nearby galaxies allow us the maximum visibility into star formation processes,
which now occur presumably under very different densities and radiation fields
than at higher redshifts.  NGST and SPECS are complementary, and both are
necessary for a complete picture.  {\bf In the Milky Way and more than a hundred
nearby galaxies, SPECS allows us to see inside dusty clouds where stars and
planets are forming.  The high angular resolution provides a new dimension of
information.}  Studies of the carbon, water, nitrogen, and oxygen lines show the
cooling of the warm phase of the interstellar gas at high spatial resolution.
Protostellar disks in local star forming complexes can be resolved in these
lines, enabling the development of more detailed astrophysical models of
protostars.  
\end{itemize}

In cosmology, the far IR and submillimeter region is centrally important.  A
plot of the luminosity of a typical spiral galaxy (the Milky Way,
Figure~\ref{milkyway}) has just two large bumps, one from 0.2 to 2 \um\ from the
luminosity of stars, and one nearly as large from 50 - 400 \um, from the energy
absorbed and reradiated by dust.  These two terms dwarf all the others, but only
one has been widely observed; the far IR emission from most galaxies is barely
detectable with present techniques, and very few galaxies can be resolved into
parts.  Early galaxies tend to be even more profuse infrared emitters than
typical galaxies in the local universe, as illustrated here by the spectrum of a
representative starburst galaxy (Figure~\ref{starburst}).  It seems likely from
direct measurements of high redshift galaxies (Blain \etal\ 1998; Barger \etal\
1998) and from measurements of the cosmic far infrared background (Hauser \etal\
1998; Fixsen \etal\ 1998) that much of the luminosity of the early universe has
been hidden by dust absorption and is reemitted in the far infrared.  The
history and even the list of constituents of the universe is presently quite
incomplete.  There is a huge uncharted far IR universe awaiting the first
discoverers with adequately sensitive instruments.

\twopanelplot{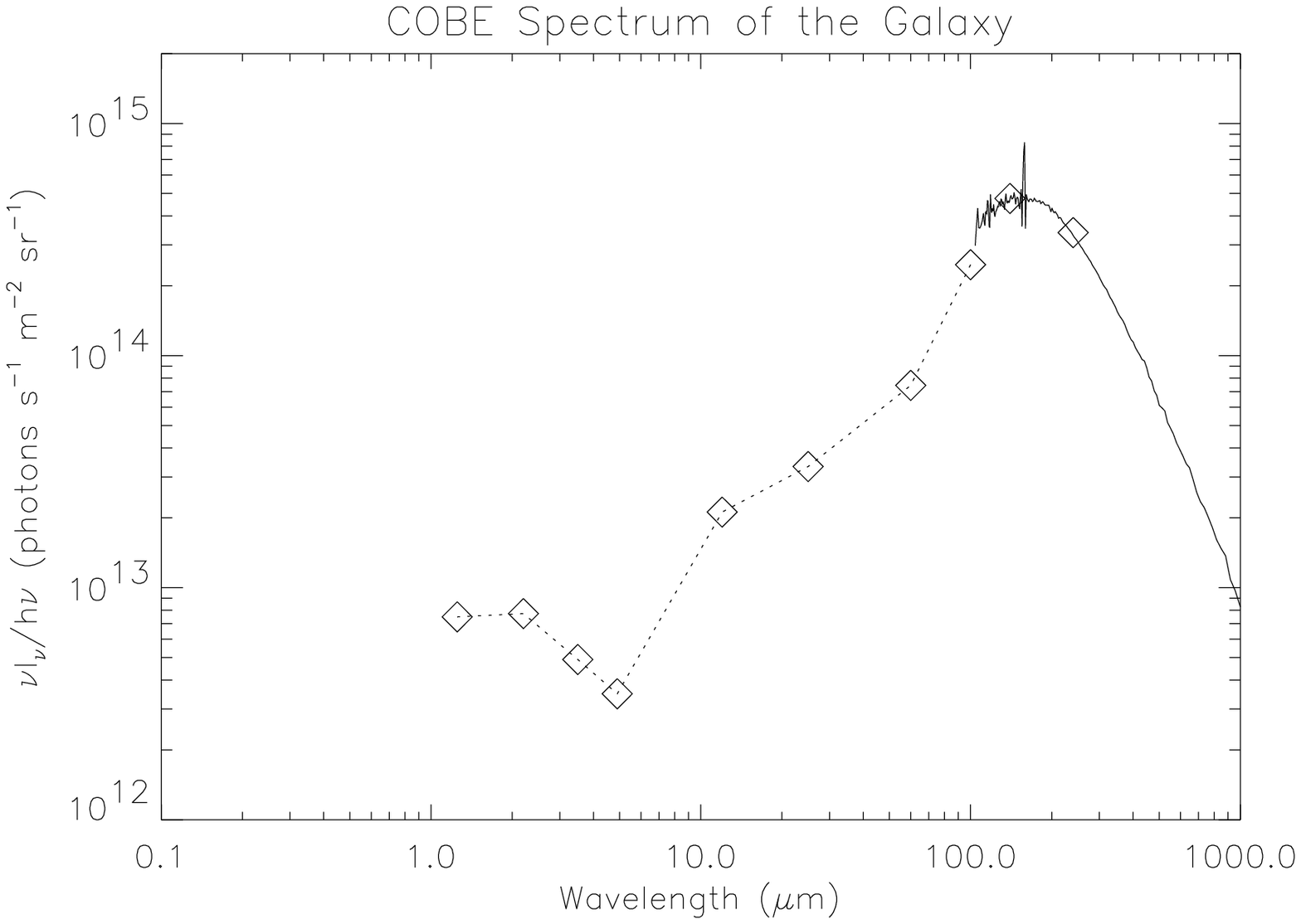}
{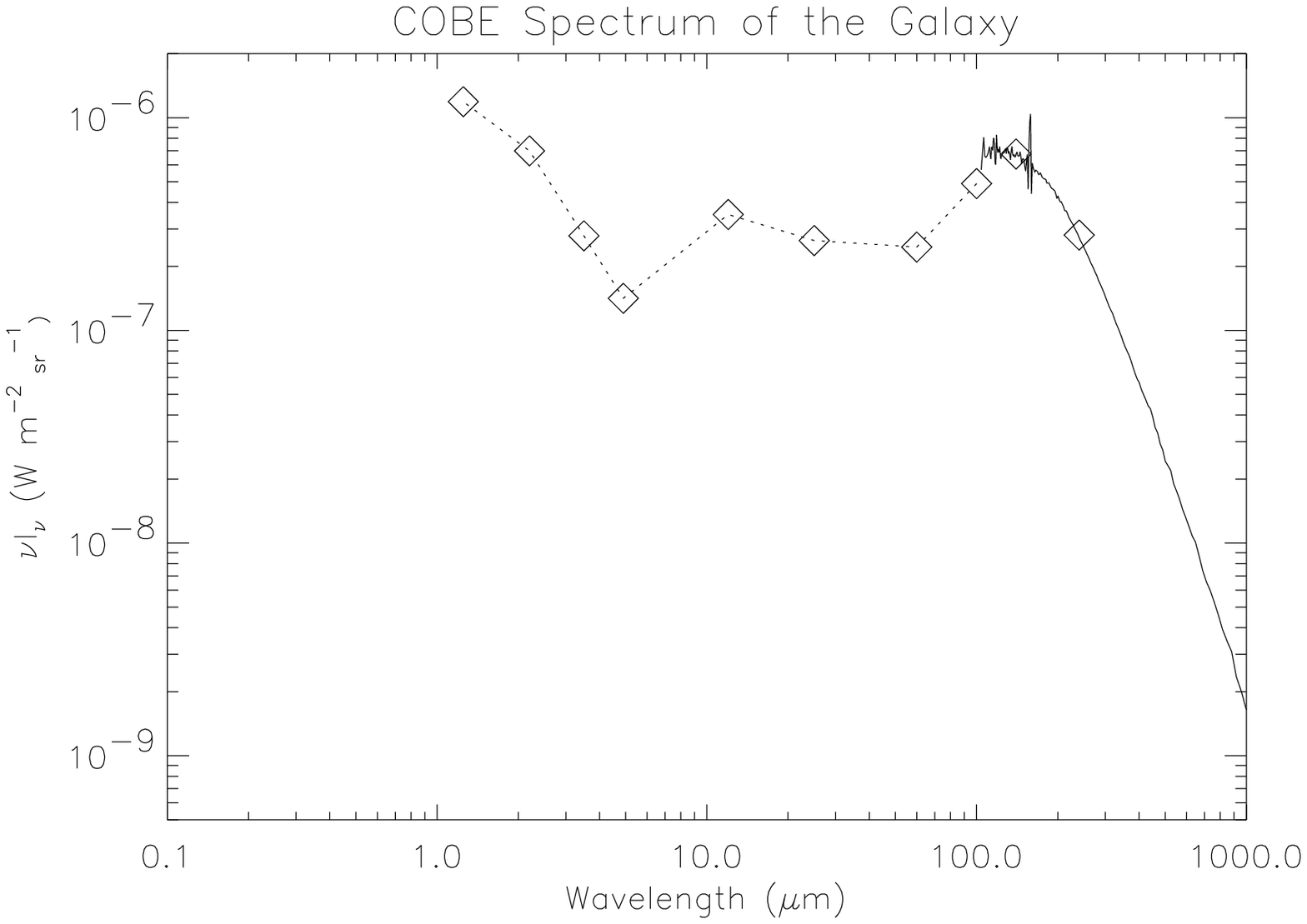}
{Milky Way spectrum in photon units (top) and $\nu$I$_\nu$ units (bottom).
Spectral lines at $\lambda \, > \, 100 \, \um$ are shown with 2\% spectral resolution. 
The C$^+$ 158 $\mu$m line, 
which cools the interstellar medium, is the brightest. Most of the 
Galaxy's photons are in the far-IR.}
{milkyway}
{0.8}

\figplot{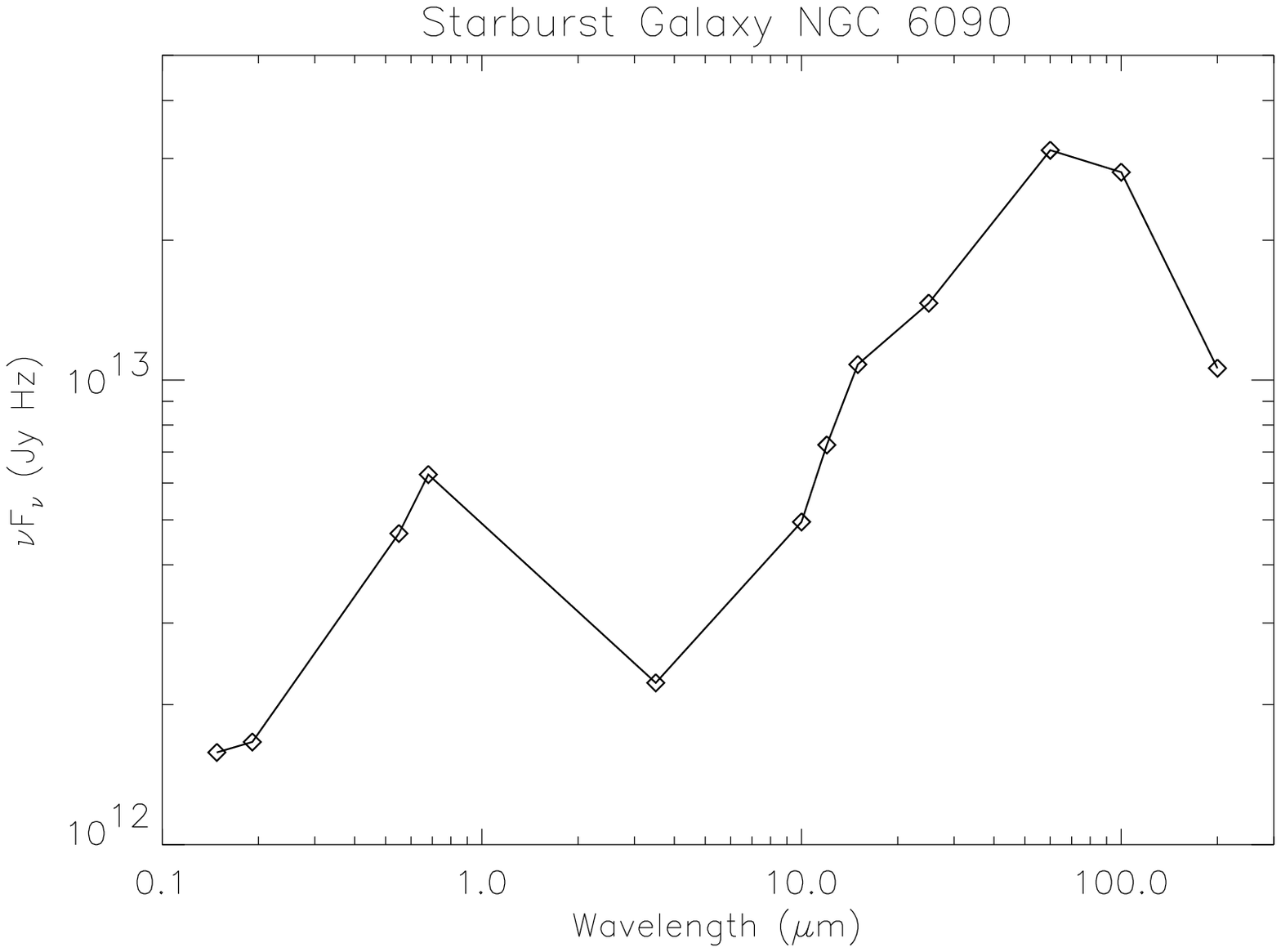}
{Starburst galaxies like NGC~6090  
are much more luminous in the far-IR than at UV and visible wavelengths, and
the dominance of the far-IR bump in photon units is even greater. The NGC~6090 
data sources are Acosta-Pulido \etal\ 1998 (ISO), Soifer \etal\ 1989 (IRAS),
and the NASA Extragalactic Database (UV - visible).}
{starburst}
{1.0}

To summarize, an instrument like SPECS is needed to fill a gap in observational
capability at far infrared and submillimeter wavelengths that translates
directly into an inability to solve some of the greatest mysteries currently
recognized in astronomy.  The SPECS observations would be complementary to those
of the NGST and MMA.  A triad of observatories consisting of these three would
allow us to take the next leap forward in our understanding of the universe.

\section{Physical Processes in Submillimeter Emission}
\label{sec:physproc}

The purpose of this section is to relate the observational capabilities
envisioned for SPECS (Table~\ref{tbl:parameters}) to the primary scientific
objectives (Table~\ref{tbl:goals}).  To make this connection one must consider
the physical processes that give rise to the observable emission.

The main processes known to produce far infrared radiation are thermal emission
from interstellar dust, fine structure line emission from interstellar atoms and
ions, rotational line emission from molecules, and synchrotron and
bremsstrahlung radiation from hot electrons in dense regions like active
galactic nuclei.  Atomic and molecular features that play key roles in the
collapse of star-forming clouds and in the energy balance of the interstellar
medium are unique to the far IR.  The line emission allows detailed
investigation of physical conditions such as temperature, density, chemical
composition, and ionization state.  Very small dust grains emit ``PAH features''
and nonthermal radiation as they spin at many GHz rates.  Condensed objects
(stars, planets, asteroids, and comets) all emit at far infrared wavelengths,
and if they are cold enough ($< 40$ K) they have no other emission.  In
addition, rest frame near- and mid-IR emission from high-z galaxies appears in
the far infrared, making this the relevant spectral window for observations of
hydrogen Brackett line and cool star emission from such galaxies.  {\bf A wide
range of phenomena are uniquely well observed in the far IR.}

\subsection{Structure in the Early, Metal-free Universe}

Since the heavy elements responsible for the cooling and collapse of gas clouds
in the present universe are not available for the formation of the first
generation of stars and galaxies, the detailed physical mechanisms of initial
formation must be very different.  SPECS will provide us with a unique view of
the first galaxies and protogalactic systems.  For example, assuming that
molecular hydrogen clouds were the progenitors of the first stars, the
redshifted H$_2$ 17 and 28 \um\ rotational lines could be observable.  SPECS
will resolve a hypothetical $z = 10$ object at the 100 pc scale in the 17 \um\
line.  If stars formed as early as $z = 20$, then SPECS would detect the
Br$\alpha$ and Br$\gamma$ lines at the redshifted wavelengths 85 \um\ and 45 \um,
respectively.  {\bf From SPECS observations it will be possible to learn whether
the first stars formed in pre-existing galaxies, or if stars formed earlier in
pre-galactic clouds.}

\subsection{The History of Energy Release and Nucleosynthesis}

What produces the cosmic far infrared background?  It is usually said that the
far IR emission comes from starburst activity (e.g., Haarsma \& Partridge 1998)
and (in combination with the UV-optical emission) reveals the nucleosynthesis
history of the universe.  However, it would be more accurate to say that the far
IR background reveals the luminosity history, since we do not know whether the
luminosity comes from starbursts or from AGN.  Some have
argued, combining ISO, SCUBA, and X-ray data, that much of the X-ray and far IR
background comes from AGN sources associated with and partially obscured by
intense nuclear starbursts (Almaini, Lawrence, and Boyle 1998; Almaini \etal\
1998; Fabian \etal\ 1998; Genzel \etal\ 1997; Lutz 1998; Lilly \etal\ 1998).
The actual nature of the far IR sources detected with the SCUBA array (Hughes
\etal\ 1998; Barger \etal\ 1998) is unknown.

There are several ways to investigate such far IR objects.  Many have far IR
luminosities that exceed those at other wavelengths by an order of magnitude, so
it is clearly important to observe this dominant luminosity as well as possible.
Low spectral resolution photometry over the entire wavelength range from X-ray
to far IR is required to get the luminosity; extrapolations from wavelengths
that are conveniently observed are not really sufficient.  Those objects that
are not too heavily obscured can be seen with HST, and high resolution UV images
might reveal star clusters or compact disks.  Hard X rays ($>$ 3 keV) can
penetrate high column densities of obscuring material, directly revealing a
non-thermal source.  Mid IR (5 - 30 $\mu$m) spectroscopy (preferably spatially
resolved) can observe the 5 - 12 $\mu$m AGN/starburst diagnostic lines and dust
features found by ISO (Genzel \etal\ 1997), with the observable spectra
depending on the depth of obscuration and on the redshift.  Long wavelength high
angular resolution imaging (e.g., with the MMA or a space interferometer) is
also required, to locate the sources precisely and determine their spatial
structure.  Very high spatial resolution is required to measure the diameter of
a compact source at the core, if there is one, calling for a millimeter array or
a far IR space interferometer with very long baselines.

SPECS would fill a very important gap.  The objects seen with SCUBA are at the
milliJansky level at 850 \um, and are considerably brighter (though less easily
seen from the ground) at shorter wavelengths.  The SPECS sensitivity is of the
order of 10 $\mu$Jy (see \S \ref{sec:SPECS}), two or three orders of magnitude
below the SCUBA source brightness, and {\bf it would have the combined spatial
resolution and sensitivity to resolve the SCUBA sources into hundreds of
pixels.}  In the end, we would know how and where most of the cosmic energy in
the whole X-ray to far IR range is liberated, something we can not know today.

In normal spiral galaxies, the [C~II] line at 158 \um\ is often the brightest
emission line at any wavelength, dominating the cooling of the Warm Neutral
Medium.  In typical spirals, the line can have a total luminosity about 1/3\% of
the total system luminosity (Stacey \etal\ 1991).  Although 
a deficit is seen in the 158 \um\ line emission from ultraluminous infrared
galaxies, suggesting that perhaps high redshift galaxies show weak [C~II] 
emission (Luhman \etal\ 1998), 
the line still may prove to be a useful marker for such galaxies.  {\bf SPECS
observations of galaxies over a wide range of redshifts in the [C~II] and other
fine structure lines from oxygen and nitrogen will enable us to trace the
buildup of heavy elements in the universe over time.}  Moderate spectral
resolution ($\sim 10^3$ - $10^4$) will be needed to detect the line emission.
High angular resolution is required to examine the mechanisms responsible for
dispersing the metals in a large number of galaxies.

\subsection{The Evolution of Galaxies and Structures}

High spectral and spatial resolution observations will yield both kinematic and
morphological information about the structures in the universe.  We would like
to know how these structures interacted and developed over time into the $z = 0$
galaxies.  What were the progenitors of halos, bulges and disks, and galaxy
clusters and superclusters, and when did these structures form?  Did galaxies
form from the ``top down,'' or were they assembled from bits and pieces?  How
did the different galaxy morphological types develop?  {\bf The broad and
continuous spectral coverage provided by NGST, SPECS and the MMA, will enable us
to follow the development of structure from $z = 10$ to $z = 0$ in a fixed set
of spectral lines and features.}

\subsection{Galaxies as Astrophysical Systems}

Following the initial phase of nucleosynthesis in the universe, the dominant
cooling and diagnostic lines are in the far IR region, and are not accessible to
detailed study from the ground (except at very high redshift).  This spectral
region includes the fine structure lines of [C~II], [O~I] and [N~II], which
dominate the cooling of and reveal the physical conditions in neutral and
ionized gas clouds.  These elements are also the building blocks of life, and it
is important to know when and where they are produced and turned into planets.

Molecular lines in the far IR from CO and H$_2$O dominate the cooling of dense
molecular clouds and are thus critical players in the formation of stars.
Additional diagnostic lines, such as those of the hydrides (OH, CH, FeH, etc.)
are also found in this spectral region.  Thus spectral line measurements will
allow us to probe dense, cold material in dark clouds, providing insight into
the details of cloud collapse in star formation.  Longer wavelength studies with
ground based telescopes of other less energetically important lines have been
remarkably successful, but we are seriously limited at present by our inability
to observe the lines that dominate the energy balance.

In addition to the far IR fine structure lines mentioned above, many galaxies
are luminous sources of {\em mid} IR fine structure emissions from Ne~II (12.8
$\mu$m), O~III (52, 88 $\mu$m), Ne~III (15.6, 36.0 $\mu$m), Ne~V (15.6, 36.0
$\mu$m) and several other ions that result from photoionization by radiation
shortward of the Lyman limit (e.g., Moorwood \etal\ 1996).  Even at moderate
redshift, all of these important diagnostic lines are shifted longward of even
the longest wavelengths contemplated for NGST.  The mid IR lines provide unique
probes of the metallicity and gas density in ionized regions, as well as the
spectral shape of the ionizing radiation field (e.g., Voit 1992).  These
transitions have several important advantages over the optical wavelength lines
traditionally used to probe H~II regions:  they are not heavily extinguished by
interstellar dust; their luminosities are only weakly dependent on temperature
and therefore provide model-independent estimates of metallicity; and they
provide line ratios that are useful diagnostics of density over a wide dynamic
range.  The availability of noble gas elements (e.g., Ne, Ar) allows the
metallicity to be determined without the complicating effects of interstellar
depletion, and the availability of a wide range of ionization states (e.g., NeII
and NeV) provides an excellent discriminant between regions that are ionized by
hot stars and those that are ionized by a harder source of radiation such as an
AGN.

The spectral bump characteristic of thermal dust emission (see
Figures~\ref{milkyway} and \ref{starburst}) will be evident in galaxies dating
back to the first epoch of dust formation.  Recent observations of an individual
galaxy indicate that dust existed at z = 5.34 (Armus \etal\ 1998).  Extinction
effects become important as soon as there is dust.

{\bf SPECS will spatially resolve individual galaxies and allow us to study them
in the spectral lines that dominate the energy balance of the interstellar
medium, cloud collapse, and star formation, and in the spectral features that
signify the presence of dust.}  We can expect to observe changes in the physical
properties of the interstellar medium, and the star formation rate or process,
as a function of redshift.  It might be possible to infer, for example, the
effects of increasing metallicity on molecular cloud cooling and the rate
of star formation.  In any case, clearly, such a rich data set
would provide the basis for new and improved astrophysical models.

In summary, observations  of the early universe in the far infrared and 
submillimeter will allow us to:
\begin{enumerate}
\item Obtain a complete census of energy release in the universe as a function 
of cosmic time;

\item Probe physical conditions associated with star formation in the early
universe, including metallicity and depletion in the interstellar media of high
redshift galaxies;

\item Investigate the kinematics of the newly formed galaxies to help understand 
the role of collisions in the initiation of star formation and the growth of 
complex structure; and

\item Measure the relative importance of ionization by starlight and by
active galactic nuclei in the high redshift universe.
\end{enumerate}

Observations of star formation in the local universe are important in providing
a detailed picture of this process in environments enriched with heavy elements.

If high sensitivity and angular resolution can be extended to the far
IR/submillimeter, then new kinds of objects may be detectable.  Old planets far
from their parent stars, either in distant orbits or already escaped, could be
numerous and detectable only in the far IR.

\section{Submillimeter Astronomy Today, Opportunities for Tomorrow}
\label{sec:todaytomorrow}

To put our current submillimeter observational capability in perspective, we can
compare the present facilities with the human eye.  At a wavelength of 1 mm,
even a large telescope like the 15 meter JCMT (James Clerk Maxwell Telescope) at
Mauna Kea has an angular resolution only as good as a 7.5 mm diameter telescope
at 0.5 $\mu$m; in other words, far IR astronomy now has the angular resolution
of an ideal human eye.  Had the Hubble Deep Field been observed at comparable
resolution, most of the interesting information would have been missed.  The far
IR telescope and atmosphere are also glowing brightly, compared to the faint
objects being observed.  In one sense this is tremendously disappointing, but in
another it is a tremendous opportunity.

To meet this opportunity, several major projects are planned or being built.
The SIRTF (Space Infrared Telescope Facility), with an 0.8 meter telescope at 4
K, is now under construction.  The Far IR Submillimeter Telescope (FIRST), an
ESA Cornerstone mission with a US-contributed 3.5 meter telescope operating at a
temperature of 70 K, will fly around 2007.  The Japanese H2L2 mission is being
planned for a launch in 2012 with a 3.5 m telescope at 4 K (Matsumoto 1998).
These new telescopes represent a continuing series of more sensitive
instruments, but they will be limited by angular resolution only a little better
than the human eye.  The images of distant objects obtained with these
telescopes will be fuzzy and overlapping.  We need a far-infrared telescope that
can distinguish individual distant galaxies.  On the ground, the MMA is to be
built in Chile, an extraordinary facility with coherent receivers, upward of 40
dishes of 10 m aperture, and spacings up to kilometers apart.  This equipment
will provide excellent angular resolution, comparable to the HST.  It will be
limited in sensitivity by antenna spillover and atmospheric emission at ambient
temperature, and in dynamic range by atmospheric fluctuations, but with long
exposures will be able to find immense numbers of galaxies because of its huge
collecting areas.

The science drivers for SPECS are derived from the considerations of the
previous sections.  We have already shown how several fundamental questions
pertaining to the evolution of structure in the universe can be addressed
observationally.  {\bf None of the existing or planned facilities will provide
the combination of high sensitivity, high angular resolution, and large solid
angle at far IR wavelengths that we believe is necessary to meet the scientific
challenge.  The obvious answer is a far IR space interferometer with cold,
photon-counting array detectors.}

The development of space interferometry was recommended by the HST and Beyond
committee (Dressler \etal\ 1996) and by the Bahcall decade survey committee
(Bahcall \etal\ 1991), although the emphasis in those reports was on shorter
wavelengths.  One reason for this emphasis is the strong desire to find and
learn about planets around other stars, and the mid IR band (5 - 20 \um)
contains the key spectral signatures of photosynthetic life (ozone, water, and
carbon dioxide).  Also, far IR astronomy is so far behind other areas that its
potential can be hard to recognize.  Acting on the Bahcall and Dressler
Committee recommendations, NASA and ESA are both preparing for space
interferometers at visible and near IR wavelengths, and the technological basis
is under development.

The key new development that enables sensitive far infrared space
interferometers is the detector system.  The current generation of bolometric
detectors is already good enough to be limited by the photon fluctuation noise
of the cosmic infrared background light in wide bandwidths.  Semiconductor
bolometers can be made in arrays by two different techniques, and improvements
using superconducting thermometers and amplifiers promise to reach the extreme
sensitivities required for this mission (Irwin 1995; 
Lee \etal\ 1996, 1997; Lee \etal\
1998).  Another promising detector technology is direct detection with
superconducting tunnel junctions.  These devices convert far IR light directly
into photocurrents that can be amplified and measured.  The necessary
superconducting electron counting amplifier design has just been developed
(Schoelkopf \etal\ 1998), and there is every reason to be optimistic about
continued improvements.  For many
applications, current direct detectors are
already far more sensitive than even ideal (``quantum limited'') amplifiers or
heterodyne receivers, which suffer additional quantum fluctuation noise
equivalent to receiving one photon per second per unit bandwidth and
polarization, and can be orders of magnitude more sensitive for the SPECS
application.

Photon detectors are preferred whenever the background photon rates are small
compared with the detector bandwidths of interest.  This is true for broadband
systems at wavelengths shorter than about 1 mm, although for very high spectral
resolution the coherent receivers are competitive down to about 300 \um\
wavelength.  The sensitivity of an ideal photon detector and amplifier are shown
in Figure~\ref{neps}, using the COBE DIRBE measurements of the brightness of the
darkest part of the sky.  Note that the sensitivity worsens as the wavelength
increases, but longer observing times and slightly larger collecting areas can
compensate.

\figplot{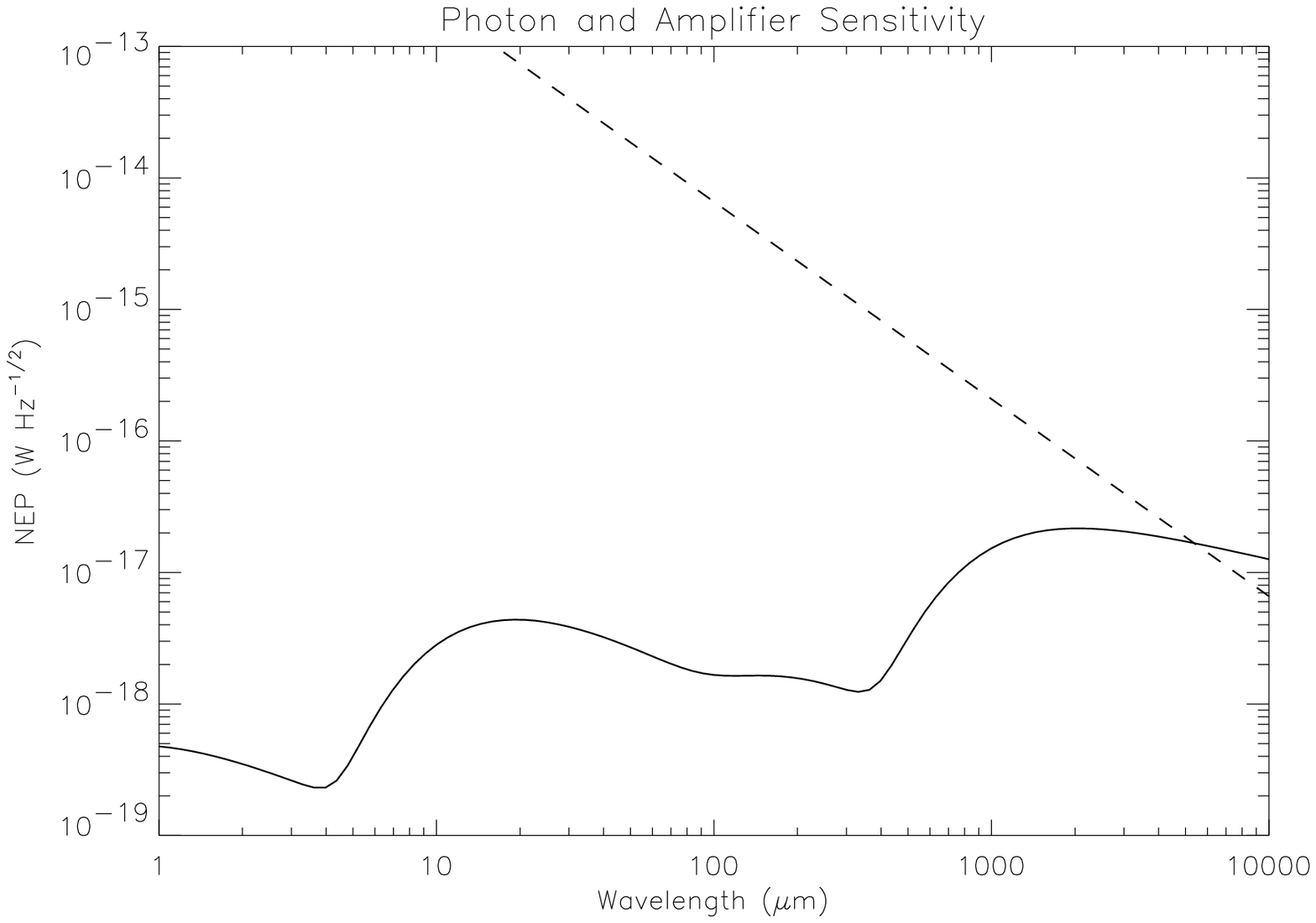}
{The sensitivity of the ideal photon noise limited detector (solid curve) and
amplifier (dashed line) for a bandwidth of 100\% and a diffraction limited
beamwidth ($A\Omega=\lambda^2$) and two polarizations.  Bolometers already
achieve sensitivities of 3 x 10$^{-18}$ W/Hz$^{1/2}$.}
{neps}
{1.0}

An interferometer would have the sensitivity of a single dish with the same
collecting area, but the angular resolution of a telescope with the full
aperture.  Considering the sensitivity improvements that are possible, it seems
that the individual apertures could be relatively small (a few meters) while the
maximum useful separation could easily be hundreds to thousands of meters.  Even
accounting for all the penalties of interferometry, such a device could map
thousands of $10^{10} \, {\rm L_\odot}$ galaxies at $z=1$ in a very reasonable
time.

\section{The Submillimeter Probe of the Evolution of Cosmic Structure (SPECS)}
\label{sec:SPECS}

The SPECS that we envision has the characteristics outlined in
Table~\ref{tbl:parameters}.  In particular, the science drivers (\S
\ref{sec:physproc}) are satisfied with several to 10 $\mu$Jy sensitivity,
several tens of milliarcsecond angular resolution, $\sim 10^4$ spectral
resolution, and a 15\arcmin\ field of view.

Below we describe the basic concepts and configurations to be considered.

\subsection{Basic Concepts}
\label{sec:concepts}

The principle of operation of a spatial interferometer is to determine
the two-point correlation  function of the incoming waves, and from that
to compute a map and spectrum. A monochromatic plane wave can be
described as $\psi(x,t)= \psi_0 \exp i({\bf \vec{k}}\cdot{\bf \vec{x}} + 
\omega t)$,
where ${\bf \vec{x}}$ and $t$ are position and time, and ${\bf \vec{k}}$ and
$\omega$ are the wavevector $(|{\bf \vec{k}}|=2\pi/\lambda)$ and frequency
$(\omega=2\pi f)$.  Its correlation function is
\begin{equation}
C({\bf \vec{x}},{\bf \vec{x}^\prime}, t, t^\prime) = E(\psi({\bf \vec{x}},t)
\psi^*({\bf \vec{x}^\prime}, t^\prime)) = |\psi_0^2| \exp i({\bf \vec{k}}\cdot
({\bf \vec{x}-\vec{x}^\prime}) + \omega( t-t^\prime)),
\end{equation}
where $E$ is the expectation value function. For a steady sky signal,
the correlation function depends only on  the spatial separation ${\bf
\vec{x}-\vec{x}^\prime}$ and the time delay $t-t^\prime$ between the
two points of observation. By custom the component of ${\bf \vec{x}-
\vec{x} ^\prime}$ perpendicular to the central line of sight is labeled
by $(u,v)\lambda$.  The component of ${\bf \vec{x}-\vec{x}^\prime}$ parallel 
to the line of sight is absorbed in the time delay term.

If there are only two telescopes at ${\bf \vec{x}}$ and ${\bf \vec{x}^\prime}$, and
the time delay $t-t^\prime$ is fixed, then the angular response function
of the correlation function is a simple cosine pattern with
angular scale of 1 cycle per
$\delta \theta = 1/ |(u,v)|$.  If the two telescopes are moved around to
measure more data points, the angular resolution of the deduced map is
approximately $\delta \theta = 1/\max |(u,v)|$. The number of
observations at different values of $(u,v)$ gives the maximum number of
parameters of the map that can be deduced from the data.  If the values
of $(u,v)$ are uniformly spaced in a square grid pattern, the algorithm
to recover the sky map is a simple Fourier transform.  Indeed, the Van
Cittert-Zernike
theorem says that the correlation function for a given ${\bf \vec{x}},
{\bf \vec{x}^\prime}$ and  an
intensity distribution on the sky is just a Fourier component

\begin{equation}
C(0) \propto \int \int I({\bf \vec{s}}) exp(-2\pi i {\bf \vec{s}}\cdot
({\bf \vec{x}-\vec{x}^\prime})/\lambda)
d\Omega,
\end{equation}
where $I({\bf \vec{s}})$ is the brightness in the direction ${\bf \vec{s}}$, 
${\bf \vec{s}}$ is a
unit vector, and $\Omega$ is solid angle. Here the $C(0)$ notation refers
to setting the
time delay equal to zero, so that the waves being correlated are in phase.

Spectral resolution can be obtained either by
choosing a narrow band filter, or by performing measurements of the
dependence of the
correlation function on the time delay.  Some spectral resolution is
required to support the spatial interferometry, because the angular
resolution and the whole scale of the reconstructed Fourier transform
image depends on wavelength.  In general the reciprocal spectral 
resolution $R^{-1} \equiv \delta \lambda / \lambda$ should be significantly 
less than the ratio of the size of the synthesized beam $\delta \theta$ to
that of the primary beam ($\lambda / D_t$), or $R > {\rm max} |(u,v)| 
\lambda / D_t$; otherwise the reconstruction is ambiguous.

We can now compare coherent and
optical technologies.  With coherent receivers, the wavefunction
$\psi$ is measured directly, amplified, and relayed to a central
computer, and all the correlation properties can be computed
electronically in a giant digital correlator.  With optical
methods, the correlation function is determined using square law
detectors (photon or energy detectors) and beamsplitters.  Assume a
beamsplitter with transmission coefficient $t$ and reflection
coefficient $r$ is used to combine two beams, one reflected and one
transmitted.  The output amplitude on one side is then
$\psi = r \psi_1 + t \psi_2$, and the intensity is $I = |r\psi_1|^2  +
|t\psi_2|^2 +  2\Re(rt^*\psi_1\psi_2^*)$, where $\Re$ is the operator
that finds the real part of a complex number.  This last term contains
the needed information about the correlation function of the two input
waves.

With coherent receivers, it is convenient to measure the correlation
function at many different time delays (lags) and to deduce the
spectrum of the incoming radiation from the Fourier transform of this
distribution. Since the waves have already been amplified, this can
all be done simultaneously in a digital correlator.  In the optical case,
the equivalent is a grating or prism spectrometer, which performs the
coherent transformations instantaneously and disperses the output photons
across a detector array.  This has the advantage that the photon
fluctuation noise is also distributed with the signal, and is the
ideal when it is possible to provide enough detectors.  Optical
technology offers another method as well, the Michelson spectrometer.
In this method, the time delay is varied by a moving mirror, and an
interferogram is measured.  The Fourier transform of this
interferogram is the desired spectrum, just as in the coherent case.
The disadvantage is that the photon fluctuation noise aflicts all the
computed frequency bins regardless of the input spectrum. On the other
hand, the Michelson spectrometer can easily cover a wide field of view
simultaneously.  This offers many potential advantages for the SPECS
mission.

The distribution of points in the $(u,v)$ plane governs the quality of
the image that can be reconstructed, and there is a large literature on
the question of ideal arrangements.  With ground based interferometers
the optimization favors a large number $m$ (10-50)  of separate antennas in a
Y shaped or circular pattern.  Then the rotation of the Earth changes the
orientation of the array relative to the sky, and each of the $m(m-1)/2$
pairs of
antennas produces a track in the $(u,v)$ plane.  Adjustable delay lines
(e.g., coax
cables) compensate for the fact that the antennas are not located in a
plane perpendicular to the line of sight.  With enough pairs and enough
Earth rotation, a large number of points can be measured in a day. To get
more coverage, the antennas can be picked up and moved, but this is time
consuming.

With a space optical interferometer, complexity grows rapidly with the
number of telescopes, so the smallest possible number is favored.  With
three telescopes, there are  phase closure relations among the
observations that enable self-calibration of certain instrument
properties (Pearson and Readhead, 1984). With four antennas, there are
amplitude closure relations as well, and with five antennas there are
enough combinations to make redundant self-calibrations.  We have not
analyzed this question fully, but are hopeful that the number of
telescopes can be kept down to three by careful use of the information
from multiple pixels within each field of view,  by careful calibration
of the equipment response functions, and by depending on good long term
stability. This is not generally done with ground based instruments
because the complexity of the receivers favors the use of more antennas
instead of more detectors.  Also, on the ground, atmospheric fluctuations
are very rapid, demanding the use of the self-calibration relations on a
second-by-second basis.  We hope this is not necessary in space.

With a space interferometer, there is a choice of methods of moving
the telescopes in the $(u,v)$ plane.  The large scale of the desired
SPECS precludes the use of rigid physical structures, so there must be
separated spacecraft flying in formation.  These can be completely
independent, or they can be tethered together to reduce fuel
consumption.  If they are tethered together, the entire collection of
telescopes can easily spin at relatively high speed around the line
of sight, producing observations of a circular set of points in the
$(u,v)$ plane. If the tethers are played in and out together, the radius
of the circle in the $(u,v)$ plane changes too, so it may be easy to
produce a spiral pattern. It would be possible to sample the $(u,v)$ 
plane completely in 2 days with 1 km maximum baselines 
if the scanning mirror could be made to stroke at about 1 Hz.
If the spacecraft are not tethered together,
then a more natural scan pattern may be a simple raster, in which
velocity changes are abrupt at the end of each sweep. Whether this is
feasible depends on the details of the propulsion system.

An advantage of a space interferometer is that in principle all spacings are
available.  There is no need to move heavy, fixed antennas from one attachment
fitting to another, an approach often used on the ground.  Elaborate algorithms
such as {\tt CLEAN} and Maximum Entropy have been developed to handle the
irregularly spaced data collected with ground-based interferometers.  Although
perfectly uniformly sampled data could be Fourier transformed directly to
produce an image, adaptations of {\tt CLEAN} and Maximum Entropy likely will be
useful in the construction of SPECS images, and might allow us to produce
high-quality images even if the $(u,v)$ plane is undersampled.  Unlimited by
atmospheric fluctuations, and with the freedom to position the light-collecting
telescopes where they are needed, a space-based interferometer can achieve much
greater dynamic range than a ground-based system.

There are important choices to be made about the spatial and spectral
resolution to be used for each target.   For unresolved sources, the
sensitivity is determined primarily by the collecting area and the
observing time, and is about the same as for a filled aperture telescope
of the same area. Excess resolution beyond what is needed for the
interesting features is a waste of observing time.  For a space
interferometer, the spatial resolution is controlled by the scan pattern
of the moving remote mirrors.  The spectral resolution should be chosen
to match either the spatial resolution, or the intrinsically interesting
spectral structure of the sources.

\subsection{Interferometer Configuration}
\label{sec:configuration}

A typical far IR imaging interferometer concept would include the following
features.  At least three telescopes are required, to take advantage of the
possibilities of self-calibration of the telescope positions with phase closure
algorithms.  They would be cooled to a low temperature, so that their thermal
emission is less than the sky brightness and they do not dominate the noise, as
shown in Figure~\ref{temperatures}.  These temperatures are now achievable with
radiation baffles and active coolers, in deep space.

\figplot{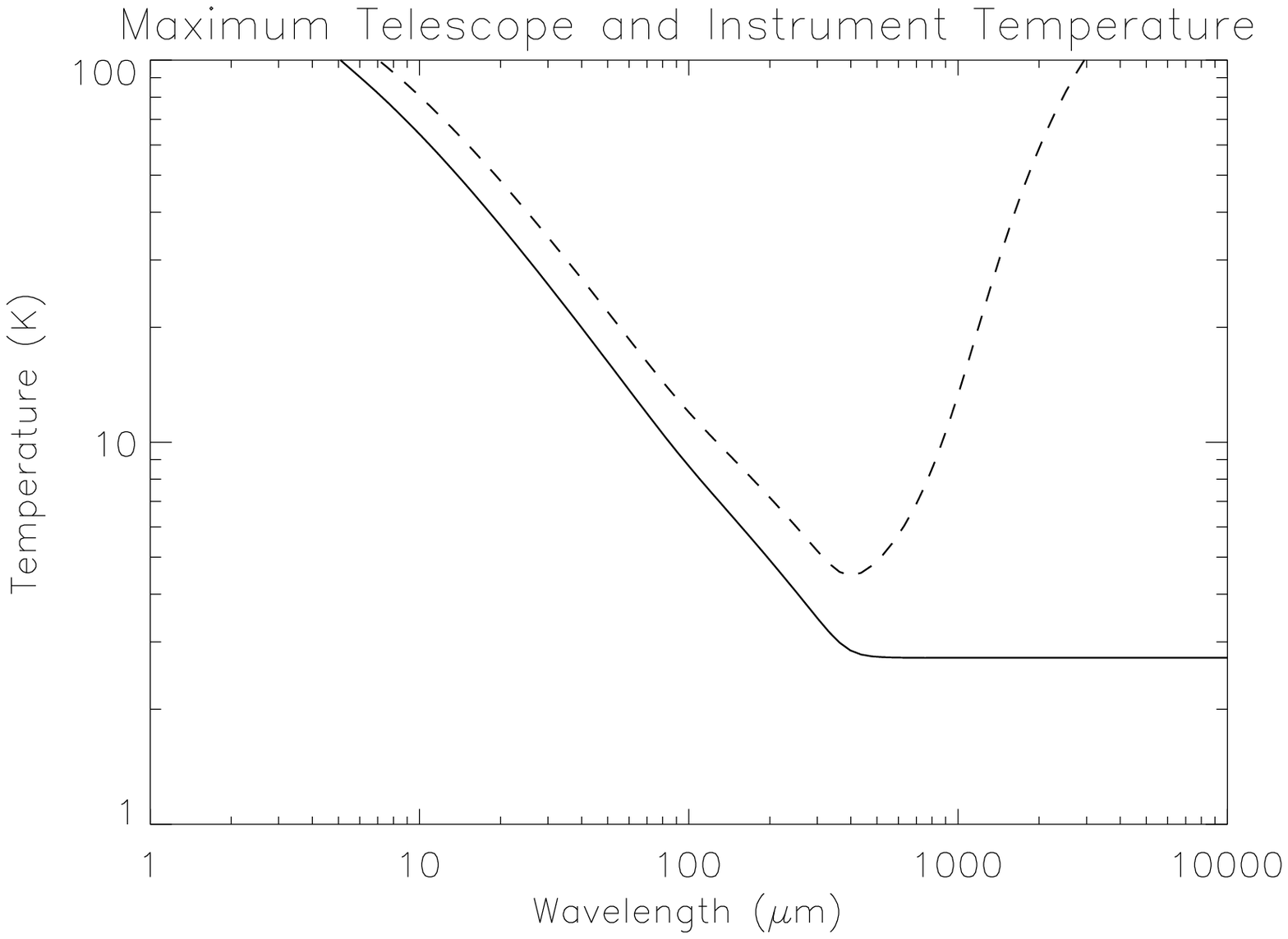}
{Required temperature for telescope (dashed curve, emissivity 0.01) and
instrument package (solid curve, unit emissivity) for thermal emission to be
less than the infrared sky brightness.}
{temperatures}
{1.0}

One way to arrange the telescopes in a far-infrared interferometer is
illustrated in Figures~\ref{telescopes-sky} and \ref{telescopes-side}.  There
are three telescopes at the central beam combining station, each looking out
sideways at a diagonal flat.  The diagonal flats move in formation to change the
location of the observing stations, and all are located equally far from their
respective telescopes.  This arrangement produces the least possible beam
divergence between the telescopes and the diagonal flats, since the telescopes
at the central station have the largest possible apertures.  It also allows the
diagonal flats to be packed close together, providing low spatial frequency
information.  The telescopes are conceived as
off-axis Cassegrain afocal systems with a magnification factor of about 10.
Their output beams reflect from small diagonal flats to become parallel to the
original line of sight from the sky, as illustrated in 
Figure~\ref{telescopes-side}.

\figplot{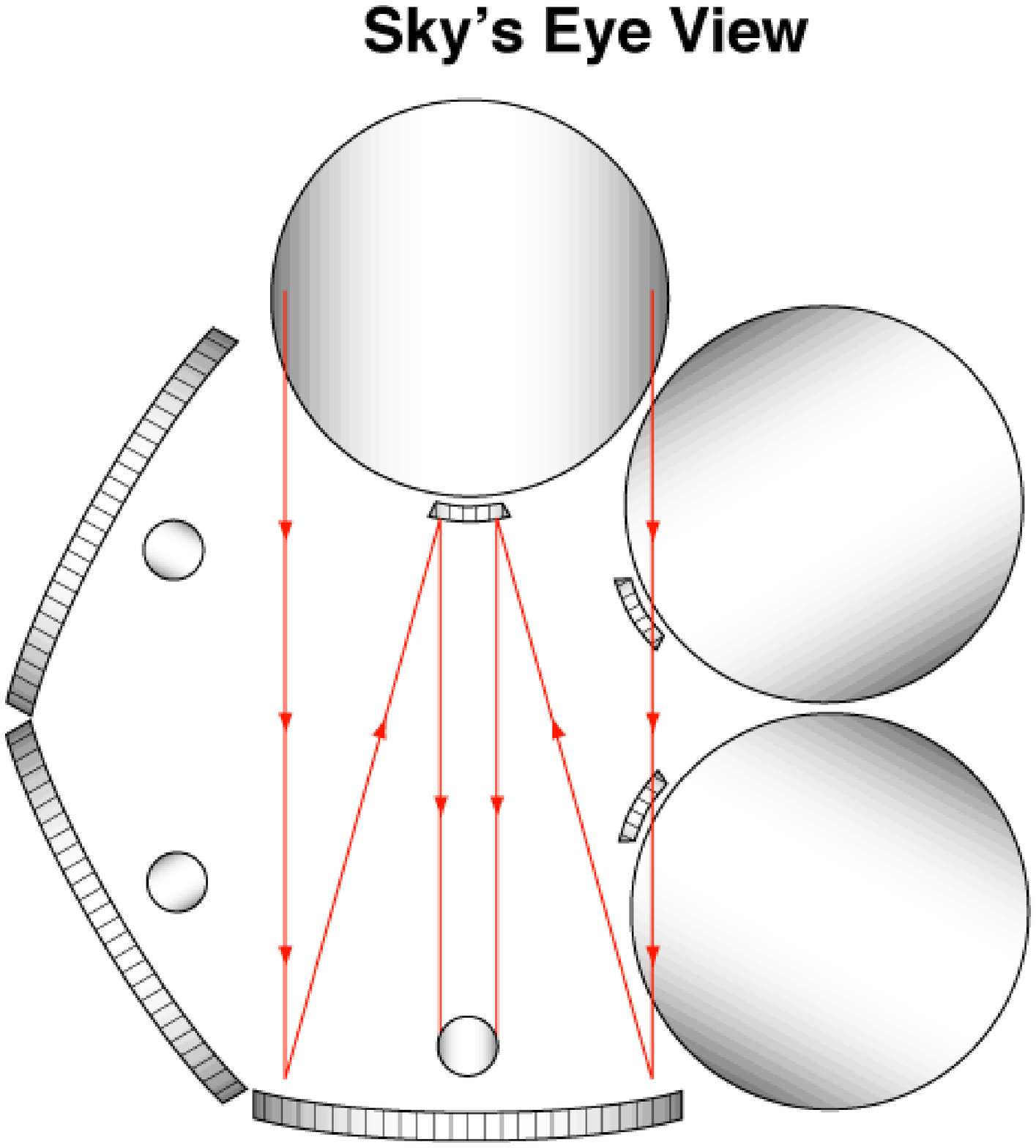}
{Telescope concept as seen from the perspective of the target of observation.
Low spatial frequency information is obtained when the mirrors are closely
spaced, as shown.  There are three Cassegrain off axis telescopes, with mirrors
seen edge on, each seeing the sky reflected from a large diagonal flat, which
appears round in this perspective.  Small circles are diagonal flats that feed
the beams down to beam combining optics.}
{telescopes-sky}
{1.0}

\figplot{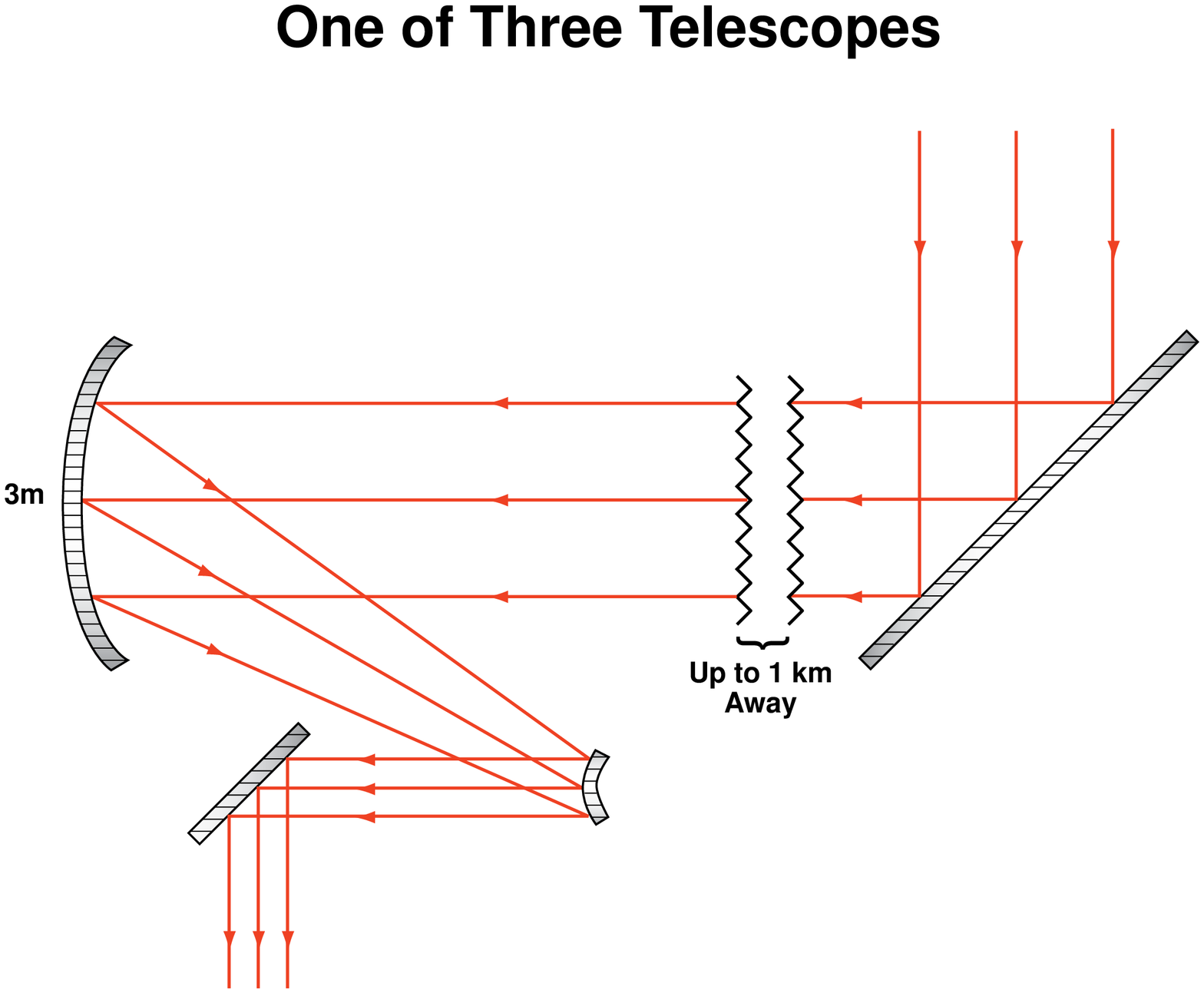}
{Telescope concept, side view. The distance between the large
flat and the primary mirror is adjustable up to about 1 km.}
{telescopes-side}
{1.0}

The telescope beams would be combined to measure the correlation functions,
using beamsplitters (as in the Michelson spectrometer) or geometrically (as in
the Michelson stellar interferometer).  We suggest the beamsplitter approach, as
that is the most similar to the successful imaging interferometers done with
microwave receivers, and the algorithms are fully developed.  In this case, the
path difference between the input beams is modulated to measure the spectrum of
the coherence function.  This is like the digital correlator for the MMA, which
determines spectra by correlating one input signal with another as a function of
time delay.  It has been shown that in the photon noise limited case, all forms
of beam combination produce approximately the same sensitivity (Prasad and
Kulkarni, 1989).  With $m$ telescopes, there are $m(m-1)/2$ beam combiners, each
with two output beams, so there are $m(m-1)$ focal planes.  One such concept is
shown in Figure~\ref{spectrometers}.  The numbers of reflections are equal in
each path, so that the images from different telescopes can be superposed
exactly.

\figplot{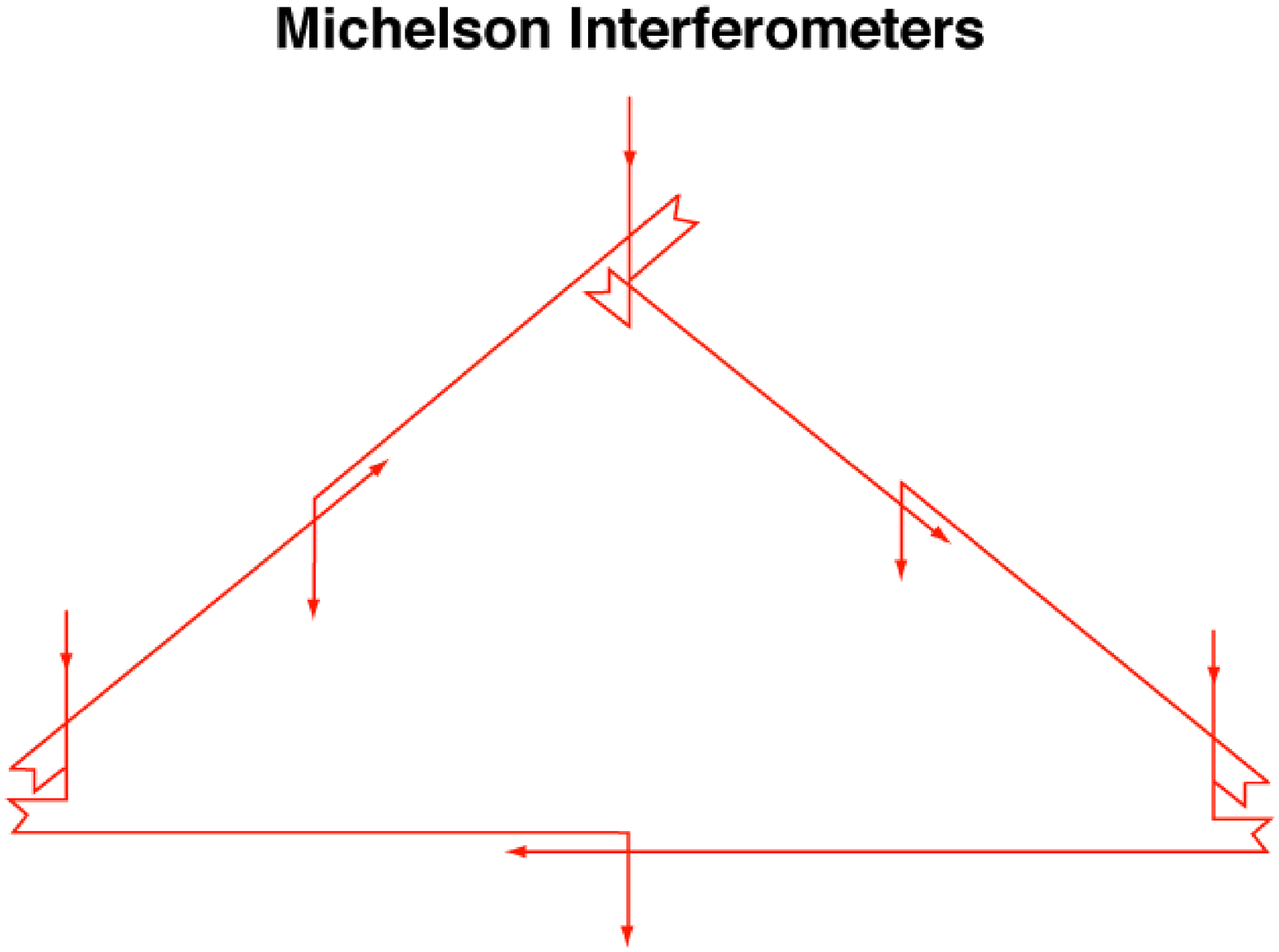}
{Beamsplitting and spectrometer concept.  The actual mirrors and beamsplitters
are omitted from the drawing, which shows only the ray paths.  Three beams come
down from the telescopes above.  Each is split into two parts at the first
vertex, a transmitted beam continuing downwards and the other reflected
sideways.  The transmitted beam strikes a reflector and also goes sideways.
Each beam is then sent to one of six Cat's Eye retroreflectors. The Cat's Eye is a
concave reflector with a secondary mirror that returns the beam parallel to
itself and sends it towards a beam combiner.  In this figure, each ``W" in the
ray path represents a Cat's Eye.  At the beam combiner in the middle of each
side of the triangle, one beam is reflected downward and then passes to a
beamsplitter, while the other beam goes directly to the beamsplitter.}
{spectrometers}
{1.0}

With the beamsplitter approach each focal plane can be a direct image of the
sky, feeding an array of detectors just as though it were at the focal point of
a single telescope.  The field of view can be as large as the telescope
aberrations and detector technology allow; we think a field of 1/4 degree may be
feasible.  This is totally different from the standard microwave array, in which
each antenna feeds a single pixel receiver.  As a result, the far IR version can
have a huge advantage in observing speed, proportional to the number of
detectors.  Effectively there are now $N_{\rm pix} \times N_{\rm pix}$ 
separate interferometers
operating simultaneously, each with its own field of view of approximately
$\Delta \theta = \lambda/D_t$, where $D_t$ is the diameter of each telescope.  Since
this $\Delta \theta$ can be much smaller than the geometrical field of view of the
telescope, it is possible to provide a large number of pixels.  This is in
addition to the sensitivity advantage of each detector over a coherent receiver.
With the beamsplitter approach, all frequencies are modulated and observed
simultaneously, much as they are in the digital correlators.  However, the
digital correlators use amplified copies of the input signals and therefore have
different noise characteristics.  Digital correlators have the advantage that
they can measure many different time delays simultaneously, partially
compensating for the lower detector sensitivity of coherent systems in low
background conditions, but they can not be used in photon background limited
receivers.

The maximum spectral resolution achievable is governed by the beam divergence at
the beam combiner, because the path difference is multiplied by the cosine of
the angle of each ray from the central ray, and a range of angles corresponds to
an apparent range of wavelengths.  With a large diameter system like that
described here, there is no problem achieving resolutions of the order of
$R=10^4$, although the path difference required becomes large, approximately
$R\lambda$.  If this much spectral resolution is required, a more compact
spectrometer such as the Fabry-Perot filter would be favored, and would have the
advantage of reducing the photon noise from signals outside the bandpass of
interest.

The spatial resolution obtained is governed by the maximum spacing of the
mirrors.  At a wavelength of 0.5 \um, the HST mirror is 4.8 million wavelengths
across.  For the same angular resolution at 200 \um, the mirrors need to be 960
meters apart.  This is clearly impossible with a rigid physical structure in
space, but it could be achieved with formation flying.  The technology to do
this will be demonstrated with the DS-3, a New Millennium program mission
scheduled for launch in 2002.  The DS-3 is intended to achieve much higher
positional accuracy and stability than are required for a far IR interferometer.
Because the remote mirrors must be propelled in a pattern, it is desirable that
their masses should be as low as possible.  For this level of accuracy, it may
be that a stretched membrane on a hoop could be made flat enough and controlled
well enough with electrostatic forces.  In any case the spacecraft engineering
challenge is significant.

The minimum spectral resolution needed is governed by the number of pixels into
which each point source may be resolved (i.e., the ratio of the primary beam
diameter to that of the synthesized beam).  Since the synthesized beam diameter
$\delta \theta$ is wavelength dependent, the wavelength must be known to a
certain tolerance.  For a general image, the spectral resolution should be
greater than the ratio of the mirror separation to the mirror diameter.

The stroke of the Michelson interferometer retroreflectors produces
interferograms that are Fourier transformed to obtain spectra of the correlation
function between two apertures.  The spectral resolution $R$ obtained is
governed by the path difference range, which must be approximately $R\lambda$
long.  In the case of the wide field interferometer described here, each
individual detector pixel has its zero path difference point located at a
different part of the mirror stroke, and the total path difference range needed
to cover the whole field of view is $\theta b_{max}$, where $\theta$ is the width of
the field and $b_{max}$ is the spacing between the remote reflectors.  Although
this is much longer than the stroke needed for a single detector pixel, by a
factor of the number of rows or columns of detectors in each array, the array is
faster than a single pixel detector in proportion to the number of rows of
elements.  At $R = 10^4$, $R \lambda \sim \theta b_{max}$.

The necessary orbit for such an interferometer must certainly be in deep space,
far from the Earth.  Otherwise, the viewing geometry relative to the Sun and
Earth is so variable that good radiative cooling is impossible, and the long
observing times required are not achievable.

The telescopes would each be shielded from the Sun by a radiative baffle like
that developed for NGST.  Additional cooling would come from an active cooler,
and at least four types are currently popular: the reverse Turbo
Brayton cycle; the Stirling cycle; the pulse tube; and the sorption pump cooler.
These coolers could reach temperatures of the order of 4 K with some effort.  The
detectors will certainly need to be colder, of the order of 0.05 to 0.5 K
depending on the technology chosen.  The additional cooling may come from
adiabatic demagnetization, as developed for ASTRO-E, or from helium-3 dilution,
as developed for FIRST.

\figplot{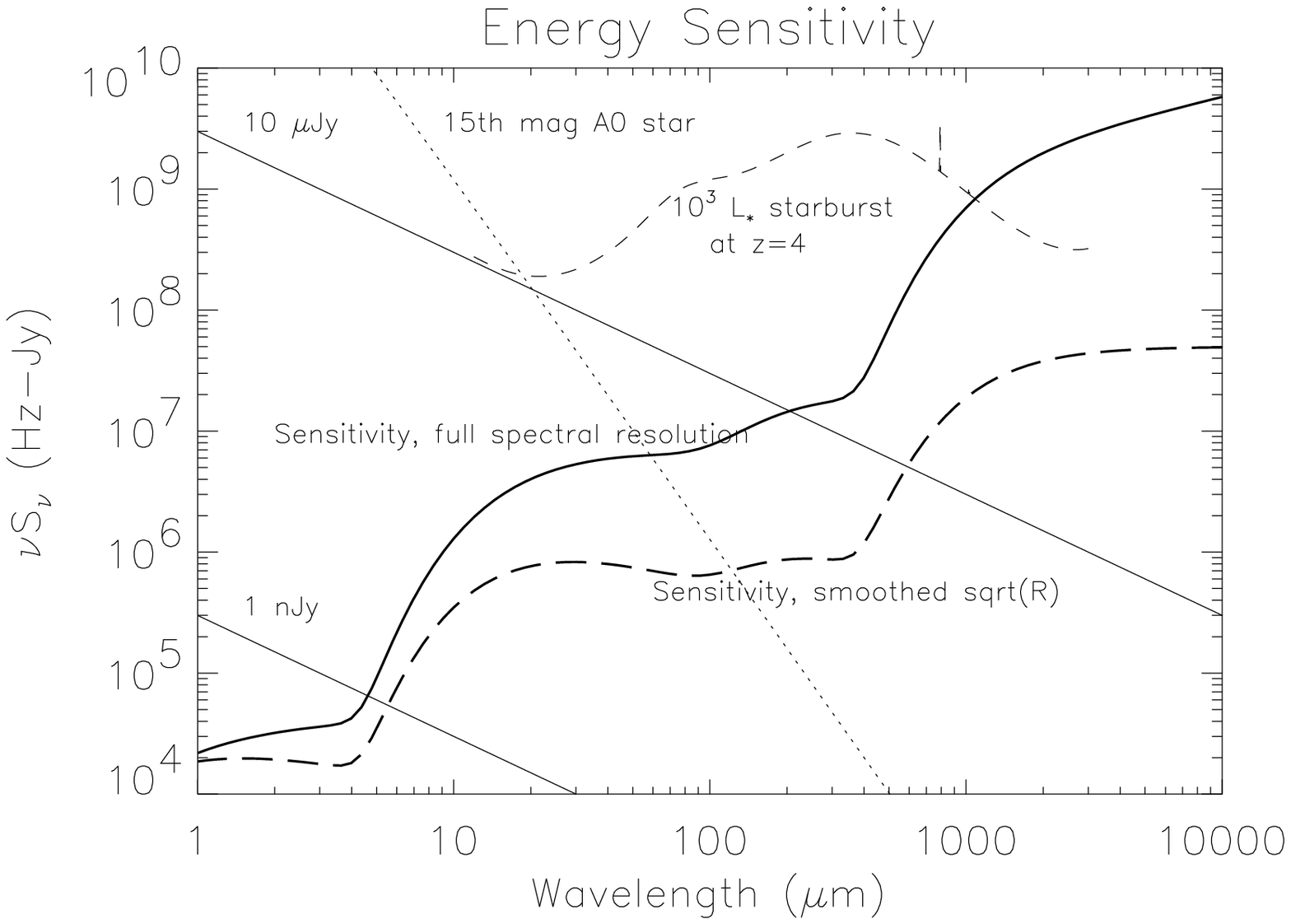}
{Sensitivity of the far-IR interferometer described in the text, both at full
spectral resolution (heavy solid curve), and with spectral smoothing (heavy
dashed curve).  The light dashed curve represents an ultraluminous (1000 L$_*$)
starburst galaxy at redshift z=4, taking NGC~6090 as a spectral template (see
Figure 1).  Such a galaxy would be readily detected by the interferometer.  The
C$^+$ 158 $\mu$m line would be redshifted to 790 $\mu$m.  The spectrum of a 15th
magnitude A0 star is shown with a dotted line.  Light solid lines mark flux
densities of 10 $\mu$Jy and 1 nJy.}
{sensitivity}
{1.0}

The sensitivity achievable by such a system is shown in
Figure~\ref{sensitivity}.  The sensitivity is $\delta (\nu S_\nu) =
{\rm NEP} \, R/(A \epsilon t^{1/2})$, where $\delta (\nu S_\nu)$ is the 
$1 \sigma$ uncertainty in the brightness, $R$ is the spectral resolution $\nu/\delta \nu =
\lambda/\delta \lambda$, ${\rm NEP}$ is the noise equivalent power from photon noise,
$\epsilon$ is the system efficiency including a sine wave modulation factor of
$1/8^{1/2}$, and $t$ is the observing time.  The ${\rm NEP}$ is for a single detector
and must be calculated allowing for the loss of photons on the way to the
detectors.  The collecting area $A$ is for the entire system, including all $m$
apertures.  The subtleties about the details of beam combination are embedded in
the efficiency factor.  For Figure~\ref{sensitivity} we have chosen three apertures of 3 m
each, an efficiency of 10\%, a background limited photon noise for a 100\%
bandwidth and that efficiency, and an observing time of 10$^5$ seconds, about 1
day.  We also assumed a spatial resolution of 0.05\arcsec , which implies a
maximum mirror separation and a required spectral resolution.  The plot shows
two curves, one with the full spectral resolution, and one with the spectra
smoothed to give a wide band image and gain a factor of R$^{1/2}$ in
sensitivity.  {\bf The sensitivity is easily adequate to reach the brightness of
a high-z galaxy, and to resolve such a galaxy into much fainter sub-units.}

At 450 \um, where Noise Equivalent Flux Density data are readily available
(Hughes and Dunlop 1997), the sensitivity of SPECS would exceed that of the
other major submillimeter facilities by factors of 50 (FIRST and the South Pole
10 m telescope), 150 (MMA), 200 (SOFIA), 700 (JCMT), or more.

We note that a smaller version of the SPECS could still have high enough
sensitivity to match the Hubble Space Telescope if the detector ${\rm NEP}$ can be improved.  This
could be feasible with new technology detectors combined with dispersive
spectrometers to limit the photon noise on each one.

\subsection{Technology Requirements}
\label{sec:technology}

Many items need to be developed to bring SPECS to fruition.  Clearly an
engineering study with a complete performance simulation is needed to define the
detailed configuration.  The major technology developments needed are:  high
sensitivity, large format arrays of detectors; cold, lightweight mirrors for
the telescopes and remote reflectors; formation flying with tethers; position measurement for
the multiple spacecraft; active coolers for telescopes and detectors; and a smooth,
long-stroke mirror scanning mechanism.  There are
many design choices to consider as well.  Narrow band filters can reduce the
spectral bandwidth and the detector noise.  Dispersive spectrometers with large
detector arrays could be combined with the imaging Michelson spectrometer to
obtain improved sensitivity.  There is no guarantee that the suggested optical
configuration is close to optimal.

\begin{table}
\begin{centering}
\begin{tabular}{|c|}
\hline
{\bf High sensitivity far IR detector arrays} \\
{\bf Lightweight, cryogenic optical systems} \\
{\bf Formation flying, possibly with tethers} \\
{\bf Active coolers} \\
{\bf Long stroke cryogenic Michelson interferometers} \\
{\bf Image processing and calibration algorithms} \\
\hline
\end{tabular}
\caption{SPECS will use new technologies, some already under development.}
\label{tbl:techs}
\end{centering}
\end{table}

There is also little practical experience with image reconstruction from
far IR detectors and spectral-spatial interferometers like SPECS.  Simple
laboratory tests should be done to wring out the problems in the analysis
algorithms; those in use at the NRAO took decades of development, and
although our requirements are similar, they are not identical.

We anticipate that the new generation of detectors will be some kind of
superconducting device, either a superconducting transition edge (TES) bolometer
with a multiplexed SQUID readout, or a superconducting junction, with photons
converted to quasiparticles that can be collected and moved to the input of a
single electron transistor.  For the junction detectors, the following problems
need to be addressed:  finding efficient devices for collecting and converting
photons to quasiparticles; transporting the quasiparticles to the amplifiers;
running the amplifiers (current designs require RF excitation); multiplexing the
detectors and electronics to enable large arrays; and building on-chip integrated
circuitry.  It is possible that an improved TES bolometer array will prove the
best choice.  In that case, developments to anticipate include:  large format
arrays, on-chip readout electronics using superconducting thermometers and SQUID
readouts, and on-chip multiplexing electronics.

\subsection{The Road Leading to SPECS}
\label{sec:roadmap}

If a concerted effort is made to advance and test the required technologies
(Table~\ref{tbl:techs}) during the next decade, it will be possible to build the
SPECS observatory in about 15 years.  A number of relevant efforts already
underway (e.g., the DS-3 mission to test formation flying) require sustained
support.  Funding for efforts that require long lead times, such as advanced
array detector and lightweight cryogenic optical system development, should be
augmented as soon as possible.  The investment needed to make the detectors a
convenient and affordable reality is vital to the future of this subject.  Early
support will also be required for SPECS tradeoff and design studies.

A likely outcome of an evaluation of cost and risk is the recommendation to fly
one or more ``precursor'' missions.  The primary purpose of such missions would
be to advance the technical frontiers in sensible increments toward the ultimate
goal, SPECS.  However, these missions would have potential scientific benefits
as well.  For example, a mission to test the detectors and cryo-coolers might
conduct an unprecedented all-sky, confusion limited far IR/submillimeter survey.
A dispersive spectrometer with a detector array could provide simultaneous
spectra.  A smaller version of the SPECS, say with 1 m mirrors, would still be
remarkably sensitive, particularly if the photon counting detector arrays work
out well and dispersive spectrometers are used to reduce the photon noise on
each detector.

Also, the US should pursue relevant opportunities to partner with other nations.
Japanese IR astronomers (Matsumoto and Okuda) have expressed interest in
contributions to their H2L2 mission (see \S \ref{sec:todaytomorrow}).  The US
could contribute, for example, improved detectors and a scanning Michelson
interferometer, which could be used to conduct a deep spectroscopic survey at R
$\sim 10^3$ (i.e., enough to reach the linewidths of typical galaxies).  While
this survey would still be confusion limited, objects at different redshifts
could be distinguished by their spectral lines.

Any opportunity that arises to test technologies applicable to SPECS on a
precursor mission required for NGST should also be exploited if it is deemed
cost-effective.

\section{Summary and Recommendations}

A space-based far infrared/submillimeter imaging interferometer is needed to
answer some of the most fundamental cosmological questions, those concerning the
development of structure in the universe (\S \ref{sec:intro}).  Such an
instrument would be complementary to the already-planned Next Generation Space
Telescope and the ground-based Millimeter Array (\S \ref{sec:context}); it would
allow access to a large number of important cooling and diagnostic spectral
lines from ions, atoms, and molecules, and to the bulk of the thermal emission
from dust clouds (\S \ref{sec:physproc}).  In light of recent technology
developments, especially the possibility of background limited photon counting
detectors in the far infrared, it is now practical to consider building the
interferometer (\S\S \ref{sec:todaytomorrow} and \ref{sec:configuration}).  The
authors recommend that we set our sights on this goal and, over the course of
the next decade, design, build and test the technology (\S \ref{sec:technology})
that will be needed to deploy a Submillimeter Probe of the Evolution of Cosmic
Structure.

\vspace{3mm}

The authors thank Chuck Bennett, Alan Kogut, Simon Radford, Ramesh Sinha
and Mark Swain for helpful comments and suggestions. We are grateful for 
and encouraged by the warm reception our SPECS concept has thus far 
received. MH and DS acknowledge funding support from NASA through
grants NAG5-3347 and NAG5-7154, respectively.

\section{References}

\frenchspacing
\setlength{\leftmargini}{0cm}
\begin{verse}
Acosta-Pulido, J. A., \etal\ 1998, A\&A, in press \hspace*{\fill} \linebreak 
  (see {\tt http://www.ipac.caltech.edu/iso/AandA/I0026.html}).\\
Almaini, O., \etal\ 1998, Astronomische Nachrichten, 319, 55.\\
Almaini, O., Lawrence, \& Boyle, B. J. 1998, COSPAR talk,
Nagoya, Japan.\\
Armus, L., Matthews, K., Neugebauer, G., \& Soifer, B. T. 1998, ApJ, 506, L89, astro-ph/9806243.\\
Bahcall, J., \etal\ 1991, {\em The Decade of Discovery in Astronomy and
  Astrophysics,} Astronomy and Astrophysics Survey Committee, National
  Academy Press \hspace*{\fill} \linebreak {\tt (http://www.nap.edu/bookstore/)}.\\
Barger, A. J., \etal\ 1998, Nature, 394, 248, astro-ph/9806317.\\
Blain, A. W., Ivison, R. J., \& Smail, I. 1998, MNRAS, 296, L29, astro-ph/9710003.\\
Dressler, A., \etal\ 1996, {\em Exploration and the Search for Origins: A Vision for
  Ultraviolet-Optical-Infrared 
  Space Astronomy,} Report of the HST \& Beyond Committee,  
  (Washington, DC: AURA).\\
Dwek, E. \etal\ 1998, ApJ, 508, 106, astro-ph/9806129.\\
Fabian, A., \etal\ 1998, MNRAS, 297, L11.\\
Fixsen, D., \etal\ 1998, ApJ, 508, 123, astro-ph/9803021.\\
Genzel, R., \etal\ 1998, ApJ, 498, 579.\\
Haarsma, D. B. \& Partridge, R. B. 1998, ApJ, 503, L5, astro-ph/9806093.\\
Harwit, M., Neufeld, D. A., Melnick, G. J. \& Kaufman, M. J. 1998, ApJ, 497, L105, astro-ph/9802346.\\ 
Hauser, M. G., \etal\ 1998, ApJ, 508, 25, astro-ph/9806167.\\
Hughes, D., \etal\ 1998, Nature, 394, 241, astro-ph/9806297.\\
Hughes, D. and Dunlop, J. 1997 ``Using new submillimetre surveys to 
  identify the evolutionary status of high-z galaxies,'' in {\em Observational 
  Cosmology with New Radio Surveys}, astro-ph/9707255.\\
Irwin, K. 1995, Ph. D. thesis, Stanford University, p. 116f.\\
Lee, A. T., \etal\ 1996, Appl. Phys. Lett. 69, 1801.\\
Lee, A. T., Lee, S.-F., Gildemeister, J. M. \& Richards, P. L. 1997, LTD-7 Conf.
Proc., p. 123.\\
Lee, S.-F., \etal\ 1998, Appl. Optics, 37, 3391.\\
Lilly, S., \etal\ 1998, for ESA Conf. Proc.
of 34th Liege International Astrophysics  Colloquium ``NGST: Science and
Technological Challenges'', astro-ph/9807261.\\
Luhman, M., \etal\ 1998, ApJ, 504, L11.\\
Lutz, D., \etal\ 1998, ApJ, 505, L103.\\
Madau, P., Pozzetti, L. \& Dickinson, M. 1998, ApJ, 498, 106.\\
Matsumoto, T. 1998, in {\em Science with the NGST,} E. P. Smith \& A. Koratkar, eds.,
  ASP Conf. Ser., v. 133, p. 63.\\
Moorwood, A. F. M., \etal\ 1996, A\&A, 315, L109.\\
O'Connell, R. W. \& McNamara, B. R. 1991, ApJ 393,597.\\
Pearson, T. J., and Readhead, A. C. S. 1984, Ann. Rev. Astron. Astrophys. 22, 97.\\
Prasad, S. \& Kulkarni, S. R. 1989, Opt. Soc. Am. J. A, 6, 1702.\\
Schoelkopf, R., \etal\ 1998, Science, 280, 1238.\\
Soifer, B. T.,  Boehmer, L., Neugebauer, G., \& Sanders, D. B. 1989, AJ, 98,
  766.\\
Stacey, G. J., \etal\ 1991, ApJ, 393, 423.\\
Voit, M. 1992, ApJ, 339, 495.\\
~\\
~\\
{\Large \bf Web site for SPECS:} \\
{\large \tt http://www.gsfc.nasa.gov/astro/specs}\\

\end{verse}
\nonfrenchspacing

\end{document}